\documentclass[prc,twocolumn,noshowpacs,preprintnumbers,amsmath,amssymb,superscriptaddress]{revtex4}

\usepackage{graphicx}

\begin{document}

\preprint{FIG12}

%%THE TITLE OF THE ARTICLE
\title{Understanding Nuclei in the upper $sd$ - shell}

%% THE AUTHORS AND AFFILIATIONS
\author{M. Saha Sarkar}
\thanks{corresponding author:  maitrayee.sahasarkar@saha.ac.in}
%\altaffiliation[Permanent address:]{Permanent address of the First Author}
%\affiliation{Affiliation of the First Author}
\author{Abhijit Bisoi}
\author{Sudatta Ray}
\author{Ritesh Kshetri}
%\affiliation{Affiliation of the Second Author}
\affiliation{Nuclear Physics Division, Saha Institute
of Nuclear Physics, Kolkata 700064, INDIA }
\author{S. Sarkar}
\affiliation{Department   of   Physics,  Bengal  Engineering  and
Science University, Shibpur, Howrah - 711103, INDIA}

%% \pacs{}  

%% THE BODY OF THE ARTICLE STARTS HERE
\begin{abstract}
Nuclei in the upper-$sd$ shell usually exhibit characteristics of spherical single particle excitations. In the recent years, employment of sophisticated techniques of gamma spectroscopy has led to observation of high spin states of several nuclei near A$\simeq$ 40. In a few of them multiparticle, multihole rotational states coexist with states of  single particle nature. We have studied a few nuclei in this mass region experimentally, using various campaigns of the Indian National Gamma Array setup. We have compared and combined our empirical observations with the large-scale shell model results  to interpret the structure of these nuclei.  Indication of population of states of large deformation has been found in our data. This gives us an opportunity to investigate the interplay of single particle and collective degrees of freedom in this mass region.
\end{abstract}
\maketitle
\section{Introduction}
Nuclei in the $sd$ shell  \cite{nndc} has been studied since 1960s. In these earlier experiments, mostly low spin states were studied. Observation 
of spherical as well as deformed states were reported. However, study of the high-spin state was scarce due to experimental limitations
in the sensitivity and efficiency of the detection system. Similarly, on the theoretical front, shell model calculations were also limited
by severe truncation. In the recent years, sophisticated techniques of gamma spectroscopy permitted observation of high spin states of 
several nuclei in this mass region. States of pure shell model characteristics have been found to coexist with those having features of permanent deformation. Superdeformation (SD) in $^{40}Ca$ \cite{40ca} and $^{36}Ar$ \cite{36ar} have been reported.
Strong violation of isospin symmetry observed in mirror pairs has been an issue of serious discussion \cite{ekman}. The nuclei in the neighbourhood
of doubly closed $^{40}Ca$ are usually  suitable for applications of spherical shell model calculations. Theoretical intrepretation of SD bands in 
upper-$sd$ shell \cite{sdth} provides an ideal opportunity to extend the microscopic description of collective rotation in nuclei
where  cross-shell correlations play an important role. Indications of breaking of a shell closure (near N = 20) far from 
stability have been observed in $^{31}Na$, $^{32}Mg$ -in the "island of inversion". Large basis cross-shell  
calculations have indicated the need for change of the $sd - fp$ energy gap for reliable reproduction 
of negative parity and high spin positive parity states in nuclei near stability also.

\begin{figure}
\vspace{5.cm}
\includegraphics{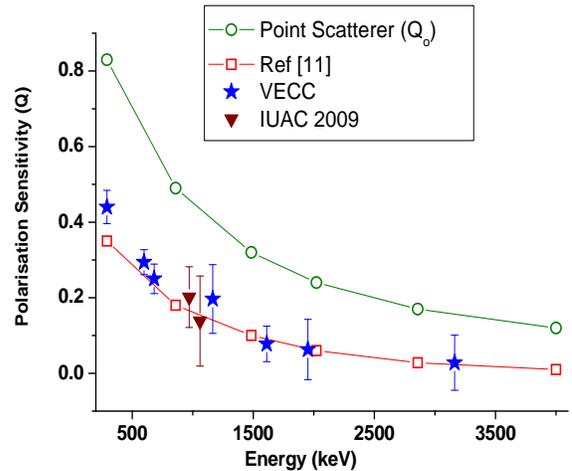}
\vspace{1.cm}
\caption{\label{polar}  
The polarisation sensitivity of a Clover
detector in different implementations of INGA.}
\end{figure}

The gamma spectroscopic studies of these light mass nuclei are different from those of the heavier ones. 
These low Z nuclei have lower Coulomb barriers. In fusion - evaporation reaction, the number of competing channels with evaporation 
of charge particles becomes very large with increasing excitation energy of the compound nucleus. At low excitation energies their level density is
also low, the energies of the gamma rays connecting these states are  usually very high ($\approx$ 2-3 MeV) where the efficiencies of the normal HPGe detectors fall off sharply. Moreover, the correlated gammas emitted  have low multiplicity, as the structure of these  nuclei are dominated by shell model states. The maximum angular momentum of the single particle orbit in this mass region is 7/2 in $1f_{7/2}$. So the angular momentum of the compound system 
is also restricted. As the spin increases, the energies of the transitions become higher implying lower detection efficiency and poorer resolution. In this respect the Clover detectors in their addback mode show excellent improvement over the normal detectors, so the Indian National Gamma Array (INGA) is an ideal setup for studying these nuclei.
 
In the present contribution, the gamma spectroscopic studies of fusion evaporation reaction residues with A$< 40$ 
done at various Indian  accelerator facilities using the INGA \cite{inga} array will  be discussed. It will encompass a few observations and related
queries which were experienced while studying these nuclei lying on or close to the line of stability.
\begin{figure}
\includegraphics[totalheight=0.4\textheight]{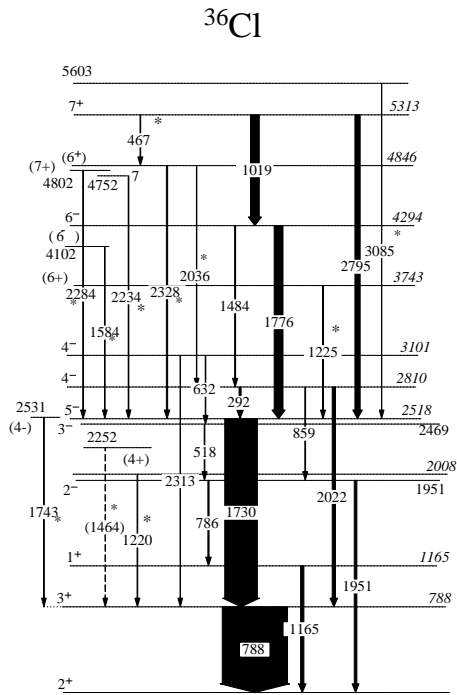}
\caption{\label{36cl}  
Experimental level scheme for $^{36}Cl$. The new transitions are marked by star(*)}
\end{figure}

\begin{table}
\caption{\label{expt} The list of experiments done at various INGA campaigns.}
\begin{tabular} {ccccc}
\hline
\multispan{2}Reaction&	Beam energy&	Centre&	No of clovers \\
Beam& Target&(MeV)\\\hline
$^{16}O $&$^{16}O$&40&TIFR&8\\
$^{14}N$&$^{27}Al$&66&TIFR&7\\
$^{16}O $&$^{27}Al$&115&VECC&8\\
$^{28}Si$&$^{12}C$&70,88,110&TIFR,IUAC, IUAC&8,8,13\\
$^{12}C$&$^{27}Al$&40&TIFR&15\\
	\hline
\end{tabular}
\end{table}

% INCLUDE ONLY EPS FIGURES
\begin{figure}
\includegraphics[totalheight=0.4\textheight]{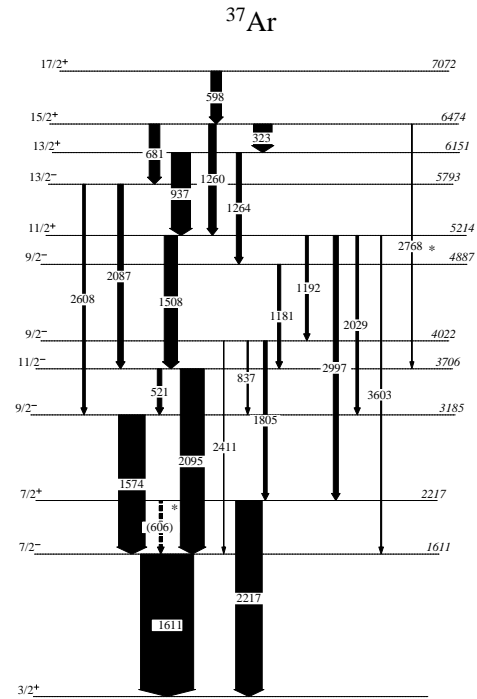}
\caption{\label{37ar}  
Experimental level scheme for $^{37}Ar$. The new transitions are marked by star(*)}
\end{figure}

\section{Experiments}
So far using different campaigns of INGA, we have been studying several upper $sd$ shell nuclei near A=40. The primary motivations for these studies are :
\begin{itemize}
\item {} To understand the issue of the variation of $sd - fp$ shell gap on the excitation spectra of nuclei on the stability line, 
{\it viz.,}  $^{35}Cl$\cite{35cl}, $^{30}P$\cite{30p} , $^{34}Cl$, $^{36}Cl$, $^{37}Ar$, etc.
\item {} In search of a 2 particle n-p state in $^{30}P$,
\item {}The issue of isospin symmetry breaking and conservation in mirror nuclei $^{35}Cl$ and $^{35}Ar$
\item {}Collectivity and deformation  in $sd$ shell nuclei like,  $^{35}Cl$, $^{34}Cl$,  $^{38}Ar$ etc.
\end{itemize}
The specifications of experiments done at different centres  to study these issues are listed in Table{\ref{expt}.
\subsection{Preludes}

As mentioned above, the gamma rays depopulating the excited states in the nuclei in this mass region usually
have very high energies ($\simeq$ 2 MeV or more). Since suitable radioactive sources having
gamma rays of energies greater than 1.5 MeV are not easily available, the energy and efficiency 
calibration of the Clover detectors were done using $^{66}Ga$ ($T_{1/2}$ = 9.41 h) along with 
standard sources like $^{60}Co$, $^{133}Ba$ and $^{152}Eu$. The radioactive $^{66}Ga$  nuclei 
emit several gamma-rays with energies ranging  from 833 to 4806 keV. This source was prepared 
by $^{52}Cr$($^{16}O,pn)^{66}Ga$ reaction at 55 MeV \cite{nima1}.
Gamma ray spectra of two (p,$\gamma$) resonances ($^{13}C (p,\gamma)^{14}N$, $^{27}Al (p,\gamma)^{28}Si$)
have been utilised  for the characterisation of the Clover detector at energies beyond
5 MeV \cite{nima2}. Apart from the efficiency and the resolution of the detector, the shapes of the full energy peaks as well as the nature of the escape
peaks which are also very crucial at higher energies have been analysed with special attention.

In all the INGA experiments, the Clover detectors are used as polarimeters for measuring polarization
asymmetry of gamma-rays to determine the electric or magnetic character of transitions. 
The sensitivity of different implementations of INGA
setup are  compared with earlier measurements of Palit et al. \cite{palit} (Fig. \ref{polar}).  
The sensitivity for an ideal point scatterer is also shown in the figure for comparison. 

\begin{figure*}[t]
\vspace{4cm}
{\includegraphics[width=2.5\columnwidth]{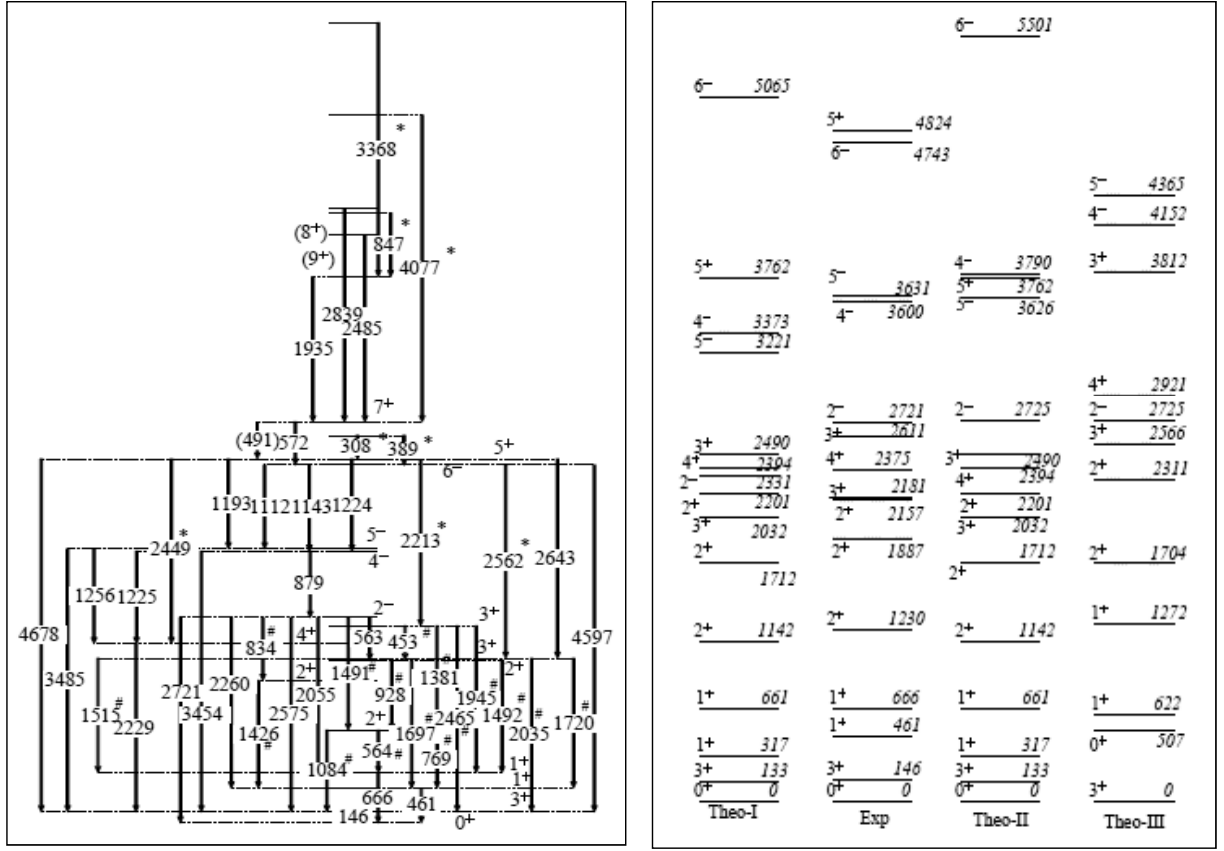}}
\vspace{-8.3cm}
\caption{\label{34cl}  
Experimental  and theoretical spectra for $^{34}Cl$. The new transitions are marked by star(*)}
\end{figure*}

\subsection{Measurements and analysis}
In most of our on-line experiments at different accelerator centres, 
we have used reactions in inverse kinematics. In  inverse kinematics, the recoils
have large velocities ($\beta$= v/c = 5-6\%) and they move within a narrow cone 
(half angle $13^o$) in the forward direction facilitating
observation of large lineshape of emitted gammas. The spread in the recoil velocities is also 
within 7\% in the inverse reaction compared to 17\% in forward reaction.
This reduction in the spread decreases the uncertainty in the lineshape fitting. 
However, in some cases,  the large shifts and broadening  pose additional problems in analysing the
lineshape data.
\begin{figure}
\vspace{2.2cm}
\includegraphics[width=1.1\columnwidth]{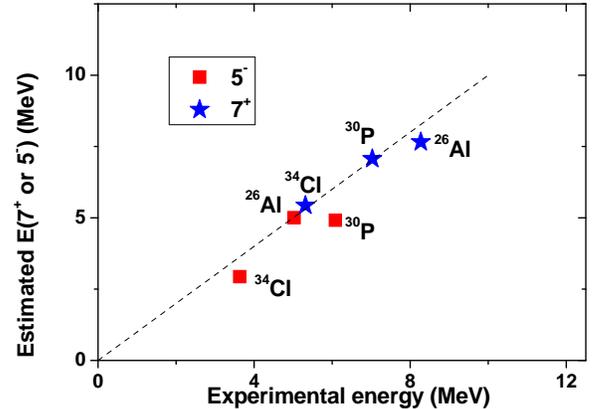}
\vspace{-4.cm}
\caption{\label{30p}  
The results of crude shell model calculations for 7$^+$ $(\pi 1f_{7/2}-\nu 1f_{7/2})$ and $5^-$ 
$(\pi 1d_{3/2}-\nu 1f_{7/2})$ states in 
odd-odd nuclei.}
\end{figure}

In these experiments, substantial number of reaction channels are populated which result in
the presence of many overlapping gamma - rays,  This makes  determination of intensities from a singles spectrum extremely difficult, if not impossible, in many cases. Therefore, relative intensities have been 
extracted from  symmetric matrices having data from all the detectors on both the axes. The multipolarities of 
transitions have been obtained through a measurement of directional correlation (DCO) of gamma -rays deexciting oriented states. For assignment of spins and gamma -ray multipole mixing ratios, the experimental
DCO ratios were compared with the theoretical ones using a standard program. 
We have performed integrated polarization asymmetry measurements (IPDCO). For this purpose two asymmetric
IPDCO matrices named parallel and perpendicular were constructed from the data \cite{35cl,30p}. 
Lifetime analysis using Doppler Shift Attenuation Method (DSAM) was done using
asymmetric   coincidence matrices having on one axis events from the detector at a particular
angle and on the second axis the coincident gamma rays registered in any other detector. Level lifetimes were extracted using both the centroid shift method and the lineshape analysis. The details about our analysis 
may be found in any one of our previous work \cite{35cl,30p}. 
%In most of the cases, the analysis program
%INGASORT \cite{rkb} was used for sorting the raw data into matrices and for generation of background
%subtracted gates from these matrices.

%% INCLUDE ONLY EPS FIGURES
%% REFERENCES START HERE
\begin{figure}
\vspace{2.2cm}
\includegraphics[width=\columnwidth]{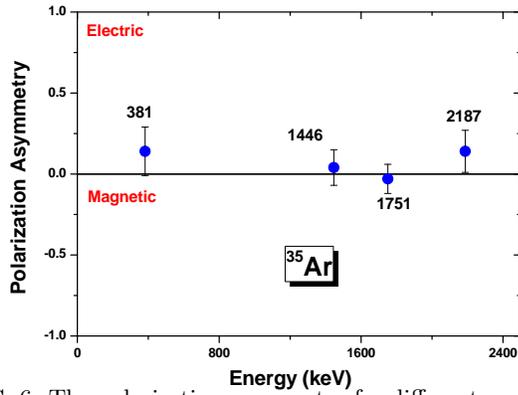}
\vspace{-4.cm}
\caption{\label{35ar}  
The polarisation asymmetry  for different gamma rays in $^{35}Ar$.}
\end{figure}

\section{Results and Discussion}
\subsection{The effect of the $sd - fp$ shell gap}
$^{35}Cl$  is a stable nucleus with  75.77\% abundance 
in natural Chlorine. With $^{16}O$ core, it has nine  and ten  valence protons and neutrons, respectively.
In its low excitation spectra, the first positive parity state at 1219 keV and the most  intense gamma is emitted 
from the lowest negative parity state at 3163 keV. Therefore, inclusion of a negative parity orbital in the valence
space is essential to explain the low energy spectrum.
In the shell model calculations, in particular for the negative parity states and for the positive parity states of 
relatively higher spins,  a nuclear Hamiltonian over the $sd - fp$  valence space is needed.
The Hamiltonian thus  consists of three parts, viz., $sd$ and $fp$ shell interactions and the cross-shell ones.
Large basis shell model calculations have been done using the code OXBASH \cite{oxb}. The valence
space consists of ($1d_{5/2}$, $1d_{3/2}$, $2s_{1/2}$, $1f_{7/2}$, $1f_{5/2}$, $2p_{3/2}$ and $2p_{1/2}$)-orbitals for both
protons and neutrons above the $^{16}O$ inert core.  The $sdpfmw$ interaction used is taken from WBMB $sd–fp$ shell Hamiltonian
\cite{wbmb}.
 Our observations from these studies are :
\begin{itemize}
\item{}To reproduce the low-lying  positive parity states \cite{35cl}, 0$\hbar \omega$ excitation has been considered, {\it i.e.}, only the full 
$sd$-shell has been used as the valence space. Low energy positive parity spectra are reproduced quite accurately with this truncation.
\item{} However, the energies of higher spin positive parity states beyond $9/2^+_1$ are predicted substantially higher than the 
experimental ones. This indicates that at relatively higher spins, the  valence space considered become inadequate for a proper description of the
state.  Nucleon excitations to the neighbouring $fp$ shell are therefore essential at these spins.
\item{} The simplest way to get negative parity states is to consider 1$\hbar\omega$ excitation, i.e. only $1p-1h$, $sd \rightarrow  fp$ 
excitations are allowed.
\item{} For these excitations, the calculated energies of the negative parity levels are consistently higher compared to the experimental values.
This feature was also observed by previous workers, {\it viz.}, \cite{35cl,others},  where the predictions for negative parity states with accuracy
better than 500-600 keV were found to be difficult. This was attributed to the overestimation of the $sd - fp$ shell gap in the corresponding 
interaction.
\end{itemize}
% INCLUDE ONLY EPS FIGURES
\begin{figure}
\vspace{3cm}
\includegraphics[width=\columnwidth]{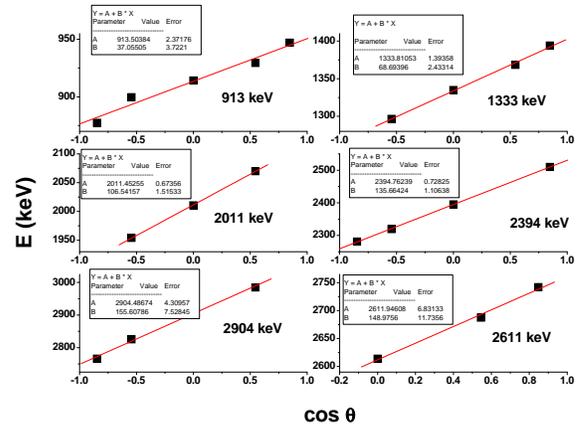}\vspace{-3.2cm}
\caption{\label{fit}  
Variation of centroid energies of shifted gammas rays with angle.}
\end{figure}
%
% 
%% INCLUDE ONLY EPS FIGURES
Later we included a few modifications in the Hamiltonian, which led to remarkable improvement of results. The observations made consequently are
\begin{itemize}
\item{}Excitations to $fp$ shell are essential to reproduce even the positive parity higher spin states.
\item{}The $sd - fp$ shell gap has to be decreased to reproduce both positive and negative parity levels. To improve the agreement for the 
negative parity  and high spin positive parities states,we have depressed the single particle energies (SPES) of $1f_{7/2}$ and $2p_{3/2}$ 
so that the first negative parity state is exactly reproduced.
\end{itemize}

We have found that  similar modifications in the $sd - fp$ shell gap are needed to reproduce our experimental data for negative parity as well as high spin positive parity  states in  $^{30}P$ \cite{30p}, 
$^{36}Cl$ (Fig. \ref{36cl}), $^{37}Ar$ (Fig. \ref{37ar}) \cite{37ar}. 
We have studied $^{34}Cl$ (Fig. \ref{34cl}), an odd-odd nucleus in the $sd$ shell, which is of interest for its importance in 
astrophysical scenario. Spherical shell model calculations have been done to interpret the experimental
data (Fig.\ref{34cl}). Several options of calculations with different values of mass normalisation for the $sd$ shell interaction as well as
variation in the monopole shift in the $fp$ interaction (Theo-I, II and III \cite{dae34cl}) have been  adopted to study the effect of the variation of
 $sd - fp$ shell gap in this cross-shell calculations \cite{dae34cl}.

\begin{figure}
\vspace{2cm}
\includegraphics[width=\columnwidth]{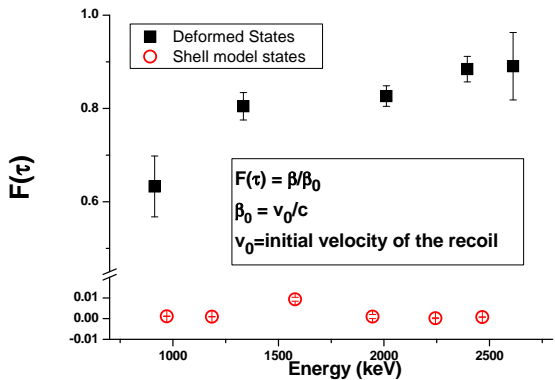}\vspace{-3cm}
\caption{\label{ftau}  
The fractional Doppler shift $F(\tau$) variation shown as a function of gamma energy. The difference between the values
for  collective and shell model states can be seen.}
\end{figure}

In a few recent work, Bender {\it et al.}\cite{34p} and Steppenbeck 
{\it et al.} \cite{30si} used the new WBP-a Hamiltonian  to calculate the negative-parity states by allowing 1p–1h excitations within a model space incorporating the $sd$ shell with the $1f_{7/2}$ and $1p_{3/2}$ orbitals. In this modified interaction, the energies of the two $fp$-shell orbits were reduced by 1.8 and 0.5 MeV to better reproduce level energies. On the other hand, Ionescu-Bujor {\it et al.}\cite{37cl} used two different interactions for this mass regions. They indicated that  the shell gap between the $sd$ and $fp$ shells produced by the first interaction ($sdfp$) is somewhat underestimated showing a need for increasing the $sd - fp$ gap, while this is overestimated for $SDPF-M$ interaction indicating a need for decreasing the $sd - fp$ gap.

\begin{figure}\vspace{2.5cm}
\includegraphics[width=\columnwidth]{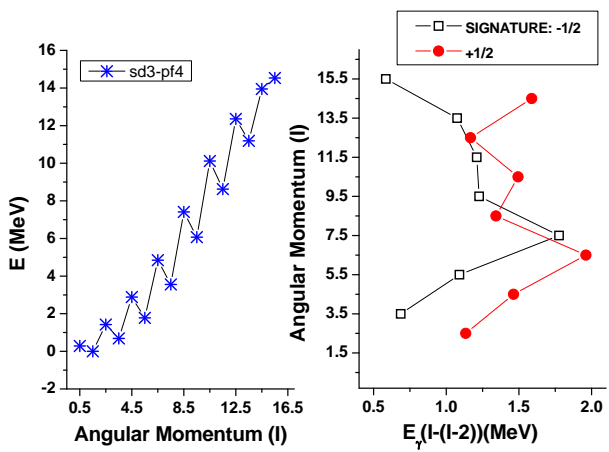}\vspace{-3cm}
\caption{\label{35cl4}  
The theoretical results for $^{35}Cl$ with four nucleons excited to the $fp$ shell. See text for details.}
\end{figure}

 To understand the origin of this difference, we carefully investigated the choices of single - particle energies in the interactions used in this mass region. There are primarily two different sets used for these calculations. In the Set A, the single-particle energies (SPE's) are determined so as 
to reproduce the neutron separation energies and the one particle spectra of $^{17}O$ ($sd$ shell) and $^{41}Ca$ ($fp$ shell).
SDPF-M, $sdpfmw$ Hamiltonians use this Set A energy. The results reported \cite{35cl,others} so far with these interactions show over-predicted energies of 
negative parity states indicating that the $1d_{3/2}-1f_{7/2}$ gap taken in this Set A may be too large. So for these interactions, 
single particle energies of $fp$ orbitals are reduced. This modification improves results for high spin positive parity 
and low spin negative parity states remarkably.

On the other hand, the $sdfp$ effective interaction \cite{37cl} takes $^{28}Si$ as a core, and the single-particle energies (SET B) are chosen in order to reproduce the
single-particle states in $^{29}Si$. The calculated energies of the negative parity states  are systematically smaller than the experimental ones indicating  that the shell gap between the $sd$ and $fp$ shells produced by the $sdfp$ interaction is somewhat underestimated. So the 
SET B is modified by increasing the $1f_{7/2}$ and $1p_{3/2}$ single-particle energies. If one analyses the experimental  energies of the first negative parity states observed in these upper-$sd$ shell nuclei, it may appear that the $sd - fp$ shell gap  evolves as a function of 
proton number \cite{daemss}. We are continuing this study for more conclusive results. This study emphasizes that the issues like erosion of shell gap and
observation of {\it "island of inversion"} thought to be relevant for nuclei away from stability, should also be studied over a broader region of nuclear territory. 
With the latest developments in the experimental facilities, the nuclei near the stability line should be re-investigated for 
finding possible correlations with those away from stability.

\begin{figure}\vspace{3cm}
\includegraphics[width=\columnwidth]{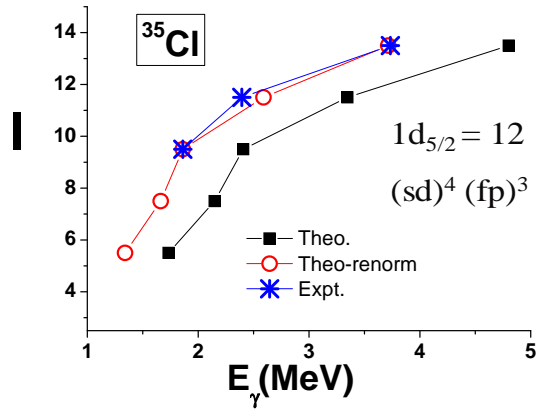}\vspace{-3cm}
\caption{\label{35cl3}  
The theoretical results  for angular momentum (I) - vs -gamma energies ($E_\gamma$) of $^{35}Cl$ with three nucleons excited to the $fp$ shell are compared with experimental data. The calculated $E_\gamma$'s normalised to the experimental data for ($19/2^- \rightarrow 15/2^-$) transition energy are also shown for comparison.}
\end{figure}

\subsection {In search of the 1$\pi-1\nu$ 7$^+$ state in $^{30}P$}
 We have studied $^{30}P$ in one of our earlier experiments \cite{30p} using the Indian National Gamma (Clover) Array (INGA) up to moderate spins (I = 5).
To understand the underlying structure of the levels and transition mechanisms, experimental data have been compared with the
results from large basis cross-shell calculations. The results for the negative parity states are especially
important in this respect. Positive parity states indicate an onset of collectivity manifested through large configuration mixing, 
whereas the negative parity states are members of  neutron-proton  multiplets. 
Later in continuation to this work, we have used the DCO   and  polarisation  measurement  data    to
determine  the  multipolarity  and  character  of  the  2858  keV transition   de-exciting   the   7202   keV  state  in  $^{30}P$.
Unambiguous assignment of spin could not be made after  analysing these data. Untruncated full $sd $ shell
calculations favour assignment of a $6^+$ option  to  the  state. 
However,  a  truncated $sd$ - $fp$ shell calculation and a comparison
of the measured and calculated half-lives of the level  from  two
calculations  do  not  totally  rule  out  a  $7^+$ opion for it. So we have done a crude shell model calculation (Fig. \ref{30p}) to identify the 
$(1f_{7/2})^2$ state by comparison with neighbouring nuclei. In this  model the $(1f_{7/2})_{7^+}^2$ states in an odd-odd nucleus (2Z+1, 2N+1)
have been obtained by exciting one proton and one neutron in $1f_{7/2}$ state. The excitation energy of that state is 
obtained by simply adding their single particle energies ($spe$).   The simplest and the best way to calculate the 
$1f_{7/2}$ proton single particle energy is from the experimental  $7/2^-$ state of the isotopic (2Z+1,2N) odd-A nucleus and for neutrons
from that in the isotonic (2Z,2N+1) odd-A nucleus.  The energy of the $(1f_{7/2})_{7^+}^2$
state in $^{30}P$ is indicated in the Fig. \ref{30p}. This study favours the assignment of 1$\pi-1\nu$ $7^+$ to the state at 7202 keV from which the 
2858 keV gamma ray in $^{30}P$ is emitted. 

\begin{figure}\vspace{3cm}
\includegraphics[width=\columnwidth]{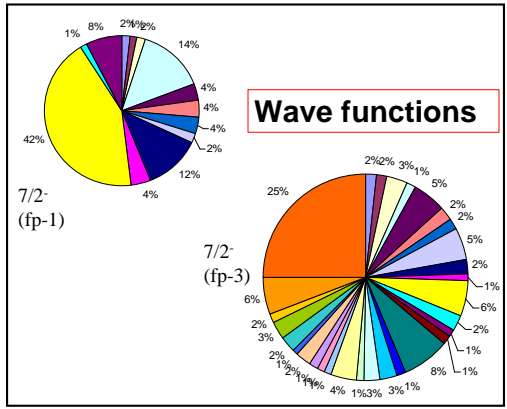}\vspace{-3cm}
\caption{\label{35clwf}  
The  wavefunctions for $7/2^-$ state in $^{35}Cl$ to demonstrate the increase in the configuration mixing with increase in
the number of nucleons excited in the $fp$ shell.}
\end{figure}

\subsection{Studies on Mirror Energy differences (MED)}
Studies on MED has been pursued to test the charge symmetry of nuclear force. It is well known that a complementary way to 
test isospin symmetry is based on investigation of the electromagnetic decay properties in mirror pairs. Anomalous MED in $sd$ shell nuclei, 
$^{35}Cl$ and $^{35}Ar$ has been observed \cite{ekman}. In the present effort, we have done measurements of the lifetimes of excited higher 
spin levels of $^{35}Cl$, mixing ratios of a few gamma transitions in $^{35}Ar$ and determined polarisation asymmetry of some gamma rays (Fig.
\ref{35ar}) of 
both the nuclei. We are analysing these data further to compare the electromagnetic decay properties of the mirror partners.

\subsection{Collectivity and deformation in $sd$-shell nuclei}
The nuclei in the neighborhood of doubly closed $^{40}Ca$ usually exhibit characteristics of single particle excitation. 
The spectroscopy of several nuclei in the mass region revealed deformed states (even Superdeformation)
at low excitation energies, indicating that the nuclei near the closed shell with Z=20  and N=20 can easily lose spherical shape.

\subsubsection{Indication of deformation in $^{35}Cl$}

\begin{figure}\vspace{3cm}
\includegraphics[width=\columnwidth]{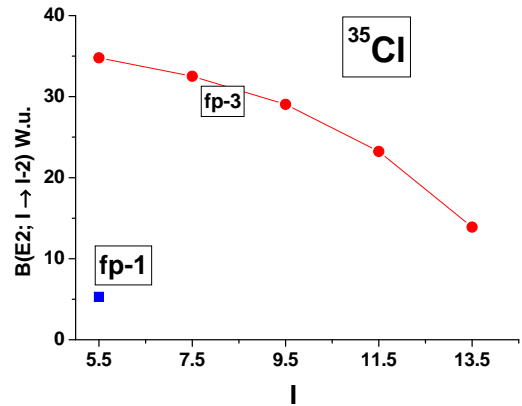}\vspace{-3cm}
\caption{\label{35clbe2}  
The theoretical B(E2) values in Weisskopf units (W.u.)  calculated for $^{35}Cl$ with one ($fp-1$) or three ($fp-3$) nucleons excited to the $fp$ shell. 
With three particle excitation, the B(E2) values show remarkable increase compared to that for one particle excitation.} 
\end{figure}

In our earlier work discussed  in Ref.\cite{35cl}, we have discussed about the six shifted peaks from around 900 keV to 3000 keV. We concluded that
the lifetimes of the corresponding states must be shorter than the characteristic stopping time of $^{35}Cl$ recoils in gold (Au) backing. 
This indicates that the corresponding transition probabilities must be large implying large deformation.
So we  investigated whether these gammas belong a single deformed band. DCO values have been extracted to determine their multipolarities. The exact energies of these shifted peaks have been determined by plotting the centroid energies as function of cos($\theta$) have been plotted (Fig. \ref{fit}). The fractional 
Doppler shift ($F(\tau))$ for these gammas are shown in Fig.\ref{ftau}. 

\begin{figure}\vspace{3cm}
\includegraphics[width=\columnwidth]{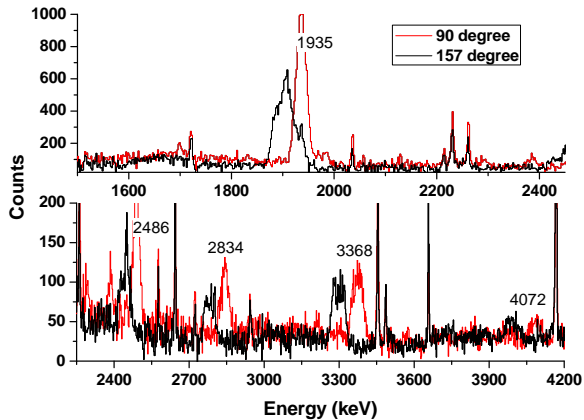}\vspace{-3cm}
\caption{\label{34cls}  
The shifted gammas in $^{34}Cl$ gamma spectrum.}
\end{figure}

 To facilitate determination of correlation between these shifted gammas a $90^0-90^0$ matrix has been generated. It has been found that only four of these 
gammas are in a sequence. All these four  gammas have E2 multipolarity. The lifetimes of these four states have been determined using Doppler shift attenuation method to get the quadrupole moment of the band.  The quadrupole moment extracted from preliminary analysis using centroid shift (F$(\tau$)) method 
 is around 0.7 eb. 

As an intial attempt, we have done theoretical calculations using a version of the Particle-Rotor Model (PRM), where experimental energies of the core can be directly given as inputs \cite{dae35cl}. Detailed large scale shell model calculations  have been done  to understand this band. 
We have done two sets of calculation to distribute 7 nucleons in the $2s_{1/2} -1d_{3/2}- fp$ orbitals. In the first set, they  are distributed with 
$(sd)^3- (fp)^4$ - as shown in Fig. \ref{35cl4}. Although the calculated states are strongly configuration mixed - they do not seem to  form a 
collective band.  The gamma energies connecting these states when plotted as a function of angular momentum show  erratic distribution for both the signatures,
instead of showing regularity. 
On the other hand, with  $(sd)^4- (fp)^3$ - a linear dependence (Fig. \ref{35cl3}) is found with a kink at 19/2$^-$. The experimental data also agree reasonably well with calculations.
A comparison between the wavefunction structures of 7/2$^-$ generated from  $(sd)^6- (fp)^1$ and $(sd)^4- (fp)^3$ (Fig. \ref{35clwf}), clearly show a strongly collective structure for $fp-3$ band. The calculated B(E2) values are also large (Fig. \ref{35clbe2}) with  gradually decreasing trend for the higher spins while approaching the termination of this band. The experimental angular correaltion, polarisation and lifetime data are being analysed in detail to reach to a firm conclusion.

For other nuclei in this mass regions, {\it viz.,} $^{34}Cl$ (Fig. \ref{34cls}) and $^{38}Ar$  gamma spectra show similar shifted peaks indicative of states with large deformation. In future we shall try to understand the evolution of collectivity  with increasing neutron number as  well as its interplay with the single particle modes in these light-mass nuclei.

\section{Conclusion}
The intruder orbitals from the $fp$ shell play an important role in the structure of nuclei
around the line of stability in the upper $sd$ shell. Experimental and theoretical studies of these nuclei near stability 
with higher precision have deeper implication for  understanding the issues relevant
for exotic nuclei in the island of inversion. Interplay between collectivity and single particle behaviour
is also important for these nuclei. Shell model calculations for explaining these experimental features  give us 
an opportunity to investigate this interplay microscopically.

\section{Acknowledgments} 
The experimental part of this work has been done in collaboration with  students and Post-doctoral fellows in the group and
our INGA collaborators. The authors would like to thank the other members of the INGA collaboration
for their cooperation.  We would like to thank the staff members of different accelerator  facilities and target laboratories
of IUAC, VECC and SINP. One of the authors
(A.Bisoi) is grateful to CSIR  for providing
the financial support.

\end{document}